\documentclass[final,twocolumn]{svjour3}
\usepackage[T1]{fontenc}
\usepackage[latin9]{inputenc}
\usepackage{listings}
\usepackage{float}
\usepackage{url}
\usepackage{amsmath}
\usepackage{graphicx}
\usepackage{esint}
\usepackage[authoryear]{natbib}

\makeatletter

\floatstyle{ruled}
\newfloat{algorithm}{tbp}{loa}
\providecommand{\algorithmname}{Algorithm}
\floatname{algorithm}{\protect\algorithmname}

\makeatother

\usepackage[british]{babel}
\begin{document}
\newcommand{\mnras}{MNRAS}
\newcommand{\apjl}{ApJ}
\newcommand{\mdash}{-}
\journalname{Statistics and Computing}
\institute{Max Planck Institut f\"ur Extraterrestrische Physik Giessenbachstrasse, 85748 Garching Germany\\johannes.buchner.acad@gmx.com}

\title{A statistical test for Nested Sampling algorithms}

\author{Johannes Buchner}

\date{10th July 2014}
\maketitle
\begin{abstract}
Nested sampling is an iterative integration procedure that shrinks
the prior volume towards higher likelihoods by removing a \textquotedbl{}live\textquotedbl{}
point at a time. A replacement point is drawn uniformly from the prior
above an ever-increasing likelihood threshold. Thus, the problem of
drawing from a space above a certain likelihood value arises naturally
in nested sampling, making algorithms that solve this problem a key
ingredient to the nested sampling framework. If the drawn points are
distributed uniformly, the removal of a point shrinks the volume in
a well-understood way, and the integration of nested sampling is unbiased.
In this work, I develop a statistical test to check whether this is
the case. This \textquotedbl{}Shrinkage Test\textquotedbl{} is useful
to verify nested sampling algorithms in a controlled environment.
I apply the shrinkage test to a test-problem, and show that some existing
algorithms fail to pass it due to over-optimisation. I then demonstrate
that a simple algorithm can be constructed which is robust against
this type of problem. This RADFRIENDS algorithm is, however, inefficient
in comparison to MULTINEST. 
\end{abstract}

\section{Introduction to Nested Sampling\label{sec:Introduction}}

For Bayesian model comparison, the key quantity of interest is the
marginal likelihood, 
\[
Z=\int{\cal L}(\theta)\cdot p(\theta)\, d\theta.
\]
It is the integral of the likelihood function ${\cal L}$ over a parameter
space whose measure is given by the prior. The nested sampling integration
framework \citep{Skilling2004} computes this integral for problems.
The strength of nested sampling not only lies with high-dimensional
integration, but also peculiar and multi-modal likelihood function
shapes can be readily handled, which pose difficulty for other approaches.
Nested sampling integrates by tracking how the part of the prior volume
reduces that is above a likelihood threshold. Like with the layers
of a Mayan pyramid, the reduction in area in a step, multiplied by
the current step height will approximate the total volume inside by
summation, regardless of the shape of each layer. The novelty is in
how the shrinking of the prior volume is tracked.

For mathematical simplicity, I will consider the unit hypercube as
the (initial) prior volume. Other priors can be mapped using the inverse
of the cumulative prior distribution, allowing broad applicability
in practice.

For one-dimensional analogy of the prior shrinkage method of nested
sampling, consider the unit interval as the prior volume. If the interval
is populated randomly uniformly by $N$ points, than the space $S$
below the lowest point is given by order statistics of order $N$
via the $\beta$ distribution: $S\sim\mathrm{Beta}(N,\,1)$, or $p(S)=N\cdot(1-S)^{N-1}$,
with the expectation value $\left\langle S\right\rangle =\left(N+1\right)^{-1}$%
\footnote{\citet{Skilling2004} uses the estimator $\left\langle \ln S\right\rangle =-1/N$,
which is better behaved at small N. For this introduction the simpler,
intuitive formula is sufficient.%
}.

If the interval above this lowest point is again filled with $N$
uniformly distributed points, we are in the same situation as at the
start, with the prior volume shrinking at each step by $\left(N+1\right)^{-1}$,
until it is $\left(1-\frac{1}{N+1}\right){}^{k}$ after $k$ steps.
In this fashion, the size of the prior volume is known on average.
For multi-dimensional applicability, what is missing is a unique and
sensible definition of the ordering. Nested sampling employs the likelihood
function values for this ordering.

To summarise, the integral $Z$ is computed by 
\begin{enumerate}
\item Randomly drawing $N$ points from the parameter space. Set $k=0$.
\item Identifying the point of lowest likelihood as ${\cal L}_{k}$ and
adding its contribution (prior shrinkage volume at this step, times
${\cal L}_{k}$) to $Z$:
\[
Z\approx\sum_{k=1}^{\infty}\left(1-\frac{1}{N+1}\right){}^{k-1}\times\frac{1}{N}\times{\cal L}_{k}
\]

\item Replacing this point by a randomly drawn point subject to having a
higher likelihood value than ${\cal L}_{k}$. Increment $k$.
\end{enumerate}
Steps 2 and 3 are repeated. This sum can be bounded by a statistical
uncertainty at every iteration step and converges \citep[see][]{evans2007discussion,chopin2007comments,skilling2009nested,Chopin2010},
so that the iteration can be stopped when the desired accuracy is
reached. If the likelihood is defined via slow-to-compute numerical
models, as often the case in the physical sciences, this poses an
additional constraint on the number of likelihood evaluations.

Nested sampling hinges (step 3) on a method to randomly draw points
which exceed a minimal likelihood value. This is known as sampling
under a constrained prior, or constrained sampling for short here.
This matter is not trivial. With peculiar shapes of the likelihood
function, multi-modality or increased dimensionality, the volume of
interest is tiny, and difficult to identify and navigate. We explore
approaches and sources of errors in the following section.

\section{Constrained sampling}

Constrained sampling, i.e. drawing from the prior but above a likelihood
threshold, has been solved in two ways, which I call \emph{local steps}
and \emph{region sampling}. Both employ the fact that the $N$ ``live''
points already lie inside the relevant sub-volume, and only another
point with such properties has to be found. Here, I discuss the potential
flaws of each method.

The first method, \emph{local steps}, starts a random walk from such
a point. After a number of Metropolis steps, by which points with
lower likelihood than required are not visited, a useful independent
prior sample is obtained. This is only the case if enough steps are
made, such that the random walk can reach all of the relevant volume.
But if the local proposal distribution is concentrated, and few steps
are made, only the neighbouring volume of the start point is sampled.
A test for detecting such a condition would be to observe the distance
between end point and existing live points. In a limited number of
geometrically simple problems, the distribution of distance to nearest
neighbour (under uniform sampling) is known, such that a constrained
sampling algorithm can be checked for correctness under such a constructed
problem. An additional limitation is that distance metrics become
less useful in higher dimensions. In practise, I have found that such
a test is less sensitive than the one presented below.

Examples of this constrained sampling approach are Markov Chain Monte
Carlo (MCMC) with a Gaussian proposal, Hamiltonian Constrained Nested
Sampling and its special approximating case Galilean Nested Sampling,
and Slice sampling \citep[see ][respectively]{Skilling2004,Betancourt2011,Skilling2012,Aitken2013}.

The second method for solving constrained sampling, \emph{region sampling},
is to guess where the permitted region lies, and draw from the prior
directly. Such a guess is augmented by the live points, which trace
out the likelihood constraint contour. The most well-known algorithm
for such an approach is MULTINEST \citep{Feroz2008,Feroz2009,Feroz2013}.
Using a clustering algorithm, MULTINEST encapsulates the live points
in a number of hyperellipses, and draws only inside these regions.
Aside from a long list of successful applications of the MULTINEST
algorithm in particle physics, cosmology and astronomy, a single problematic
case has been discovered in \citet{Beaujean2013} and analysed in
\citet{Feroz2013}. Under this perhaps pathological, but physics-motivated
likelihood definition, the MULTINEST algorithm consistently gives
incorrect results. What then can be sources of such a problem?

When constructing the sampling region, two errors can be made. The
sampling region may contain space that falls below the likelihood
threshold. This results in sampled points that are not useful and
have to be rejected. This rejection sampling affects the number of
likelihood function evaluations. In high-dimensional problems, the
spaces grow quickly, such that the fraction of useless points can
become prohibitive. In practice, the MULTINEST algorithm works inefficiently
beyond $\sim20$ dimensions \citep{Feroz2008}. However, contrary
to the ``local steps'' method above, the points obtained are guaranteed
to be drawn uniformly from the sampling region by construction.

The second and more severe type of error is the inadvertent exclusion
of relevant volume from the constructed sampling region. This under-estimation
of the prior space can lead to biased likelihood draws, either to
higher or lower values, depending on the problematic situation. To
avoid this problem, the sampling region is typically expanded by a
constant growth factor. But can such an algorithmic problem be detected,
at least in constructed test problems? I present a statistical test,
the Shrinkage Test.

\section{The Shrinkage Test\label{sec:Shrinkage-Test}}

The shrinkage of the prior volume in nested sampling is known: $1/N$
of the volume is supposed to be removed. If the shrinkage is accelerated
by inadvertently missing a sampling region, this is no longer true. 

Let us thus construct test problems where the likelihood constraint
contour is known for each removed point, as well as the volume contained.
If we compute the ratio of volumes at each step, we can compare it
to the expectation of 
\[
\left\langle t_{i}\right\rangle =\left\langle \frac{V_{i}}{V_{i+1}}\right\rangle =\frac{N}{N+1}.
\]

Any test problems can be used where the size of the constraint region,
$V_{i}$, can be computed for the current likelihood value. For instance,
for a Gaussian likelihood, the geometric volume formula of an ellipse
is applicable. But the simplest test problem is one where at each
likelihood value the contour is a hyper-rectangle. This is the case
for the ``hyper-pyramid'' likelihood function,

\[
\ln L=-\left(\sup_{i}\left|\frac{x_{i}-\frac{1}{2}}{\sigma_{i}}\right|\right)^{1/s}.
\]
Here, $s$ controls the slope of the likelihood and $\sigma_{i}$
defines the scales in each dimension. In this problem, the contours
are given directly by $L$, as $x_{i}=[r_{0}-\frac{1}{2},\, r_{0}+\frac{1}{2}]$
with $r_{0}=(-\ln L)^{s}=\sup_{i}\left|\frac{x_{i}-\frac{1}{2}}{\sigma_{i}}\right|$.
The corresponding volume is associated with a hyper-rectangle, i.e.
$V=(2\cdot r_{0})^{d}\times\prod_{i}\sigma_{i}$.

The distribution of the volume shrinkage $t_{i}=\frac{V_{i+1}}{V_{i}}$,
is given by $p(t;\, N)\sim(1-t)^{N-1}$, which can be described by
a beta distribution with the shape parameters $\alpha=N$ and $\beta=1$.
Its cumulative distribution is thus simply $t^{N}$. This function
is cornered at $R\approx1$ for reasonable values of $N$ ($\sim400$).
For nicer visualisation, lets consider the border that is being cut
away: $S=1-t^{1/d}$. The expected cumulative distribution on $S$
is then $p(<S)=1-(1-S)^{d\cdot N}$.

To test conformity with uniform sampling, the constrained sampling
algorithm is applied for many iterations (e.g. 10000). Using the sequence
of removed points, the removed volume $S$ is computed and compared
to and the expected cumulative distribution. The frequency of deviations
between the theoretical and obtained distribution can be assessed
visually. As the number of samples can be increased, discrepancies
should become clear. For quantification of the distance, e.g. the
Kolmogorov-Smirnov (KS) test can be applied.

When applying the test in this work, I will use $s=100$ and $\sigma_{i}=1$
(hyper-cube contours). However, this test can simulate a wide variety
of shapes including problems with multiple scales (e.g. with $\sigma_{i}=10^{-3i/d}$),
or Gaussian likelihoods where the contours are hyper-ellipses. The
case of multiple modes can also be considered. It should be stressed
that the dimensionality of the test can be chosen, and varied to analyse
the algorithm of interest.

\section{Application of the Shrinkage Test\label{sec:Application-shrinkage-MultiNest}}

Lets now verify whether the MULTINEST algorithm, with commonly used
parameters, passes the Shrinkage test. Other algorithms are considered
later in Section \ref{sec:Evaluation}. I use version 3.4 of the MULTINEST
library \citep{Feroz2008,Feroz2009}. I set the sampling efficiency
to $30\%$, and the maximum number of modes to 100. I use two configurations,
with 400 and 1000 live points, and without considering importance
nested sampling \citep[see][]{Feroz2013}.

\begin{figure}
\begin{centering}
\includegraphics[bb=0bp 0bp 318bp 351bp,width=0.8\columnwidth]{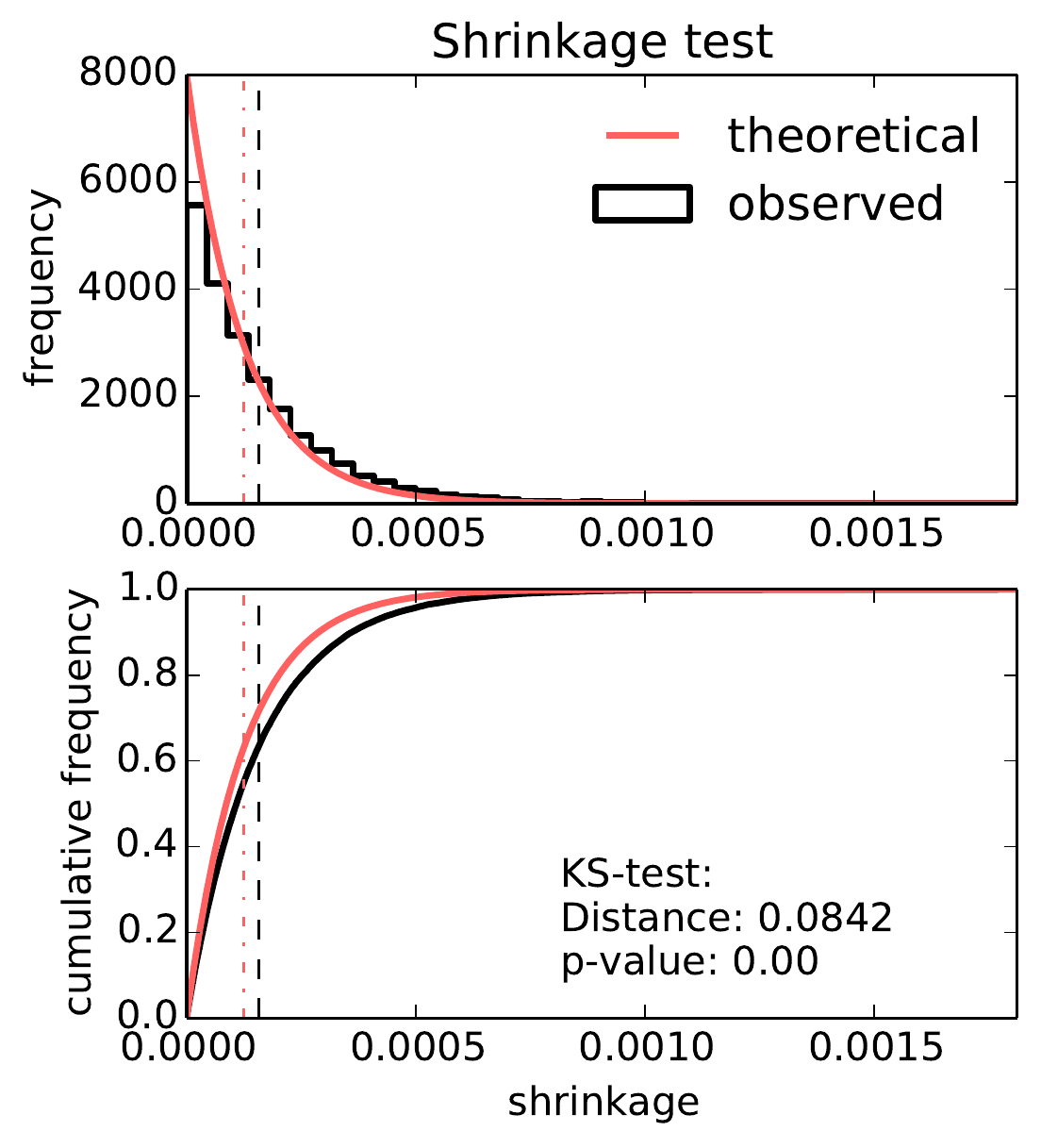}
\par\end{centering}

\caption{\label{fig:plot-multinest}Shrinkage test results. The MULTINEST algorithm
running in 20 dimensions is analysed. The panels show the distribution
of the shrinkage border (histogram in the \emph{t}\textit{op panel},
cumulative distribution in the \textit{bottom panel}). The observed
distribution (black) is shifted to higher values compared to the theoretical
distribution (red). This indicates that too much space is being cut
away. The vertical lines indicate the means of the distributions.}
\end{figure}

The Shrinkage test using the hyper-pyramid likelihood (see previous
Section) is applied. I consider 2, 7 and 20 dimensions, and run nested
sampling up to a tiny tolerance to avoid premature termination. In
each of the first 10000 iterations the newly sampled point is stored.
Using a number of such sequences, I compute the empirical distribution
of the shrinkage $S$, and plot it against the theoretical distribution.
This is shown in Figure \ref{fig:plot-multinest} for the 20-dimensional
case. I find that in 2 dimensions, the distributions match, but in
7 and 20 dimensions, the shrinkage $S$ tends to lie at higher values.
This indicates that too much space is being cut away. This test thus
shows, by discrepancy of the theoretical and real shrinkage of the
prior volume, that the MULTINEST algorithm under-estimates the volume
for this test problem, and samples from a smaller region. We have
thus identified a potential source of error relevant also for the
problem of \citet{Beaujean2013}.

\section{Robustness against accelerated shrinking\label{sec:Robustness}}

Can we then devise a rejection algorithm that does not suffer from
the problem of shrinking too quickly? Here I present an approach that
gives some correctness guarantees, but does not emphasise efficiency,
particularly in high dimensions. Here I exploit again the live points,
but also use the property that they are already uniformly distributed.
The next point ought to be in their neighbourhood too, where neighbourhood
is defined by having at most distance $R$ to a live point (this donates
the definition of the sampling region). In particular, the method
should be robust so that every live point \emph{could} be sampled
if it was not known. A initial idea is to leave each point out in
turn, compute the distance to its nearest neighbour, and use the maximum
of this quantity as $R$. Such a jackknife scheme is quite robust,
as all points are closer than $R$ to a live point. However, had the
point donating the maximum $R$ not been in the sample, it could not
be obtained. I thus go further and employ a bootstrapping-like method,
which I describe now in detail.

\section{The RADFRIENDS algorithm\label{sec:RADFRIENDS}}

\begin{algorithm}
\begin{centering}
\begin{lstlisting}[basicstyle={\ttfamily},language=C,tabsize=4]
function draw_constrained(Lmin, live_points) {
	R = compute_R(live_points)
	loop {
		p = draw_near(live_points, R)
		if (likelihood(p) > Lmin) 
           return p
	}
}
function compute_R(live_points) {
	R = 0;
	n = size of (live_points)
	for i = 1 to 50 { # bootstrapping rounds:
		chosen_set = choose n with replacement 
                     from live_points
		not_chosen_set = live_points not 
                         in chosen_set
		for each point in not_chosen_set {
			minR = shortest distance 
                   to a point in chosen_set
			if (minR > R)
				R = minR
		}
	}
	return R;
}
\end{lstlisting}

\par\end{centering}

\caption{\label{alg:RadFriends}The \textsc{RADFRIENDS} algorithm for drawing
a new sample from the prior, under the constraint that its likelihood
is larger than $L_{min}$. The \texttt{draw\_near} procedure is explained
in the text and shown in Algorithm \ref{alg:draw_near}.}
\end{algorithm}

\begin{figure*}[p]
\begin{centering}
\includegraphics[width=1\textwidth]{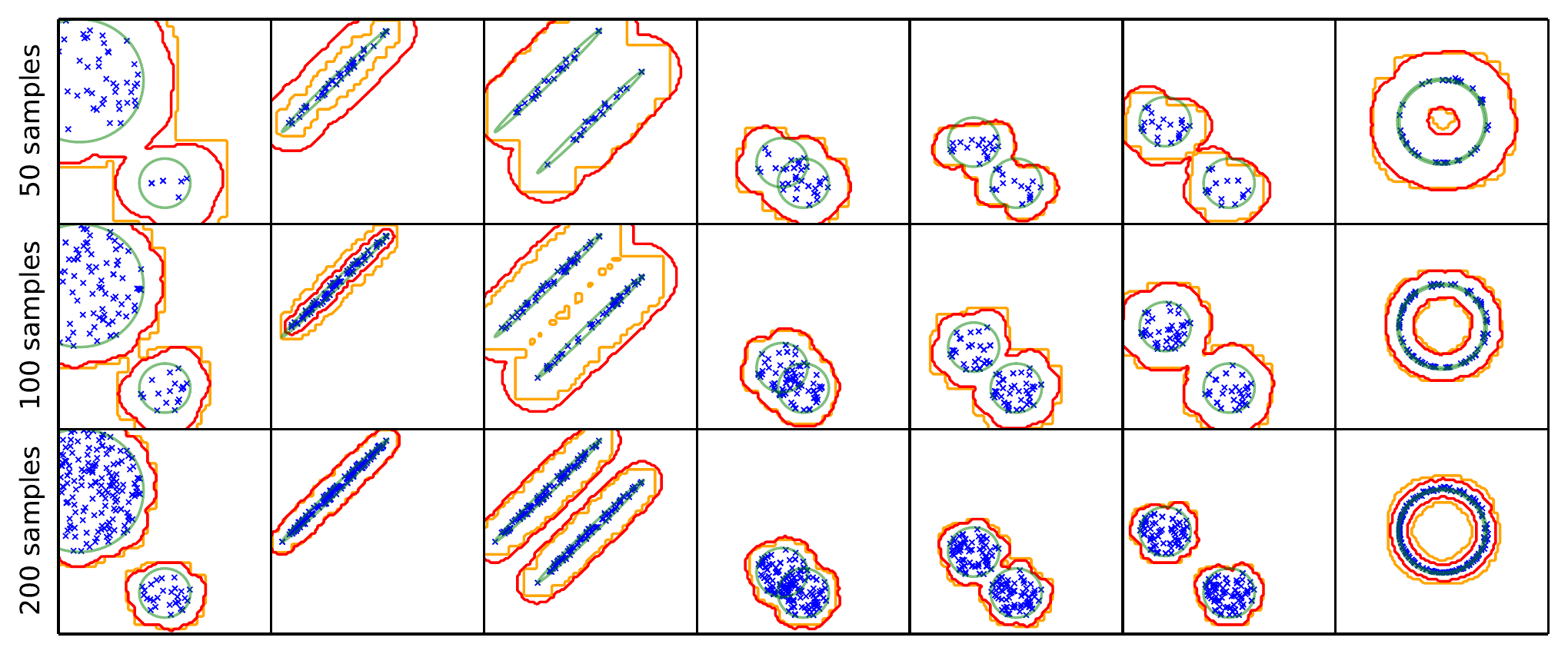}
\par\end{centering}

\caption{\label{fig:RadFriends-Contours}Examples of the sampling regions for
the\textsc{ RADFRIENDS }algorithm, after employing the \texttt{compute\_distance}
procedure. The blue crosses indicate the live points used for each
test case, which are drawn uniformly from the (in practice unknown)
likelihood constraint region (green circular lines). The sampling
region used by \texttt{draw\_constrained} is shown for a Euclidean
norm (red line) and a supremum norm (orange). From top to bottom,
the number of live points have been increased (50, 100, 200 samples).
A general trend of narrowing can be observed. These examples highlight
how the algorithm adapts to the peculiar shape of the region of interest
(e.g. second and right-most panel), and can handle multiple modes
(third to sixth panel) without any assumption on the shape.}
\end{figure*}
\begin{algorithm*}[p]
\hfill\begin{minipage}[t]{.45\textwidth}

\begin{centering}
\begin{lstlisting}[basicstyle={\ttfamily},language=C,tabsize=4]
function draw_near(live_points, R) {
    loop {  # variant 1:
      candidate = draw from unconstrained prior
      mindistance = shortest distance of 
                    candidate to a point 
                    in live_points
      if (mindistance < R)
         return candidate
    }
}
\end{lstlisting}

\par\end{centering}

\end{minipage}\hfill\begin{minipage}[t]{.45\textwidth}

\begin{centering}
\begin{lstlisting}[basicstyle={\ttfamily},language=C,tabsize=4]
function draw_near(live_points, R) {
    loop {  # variant 2:
      mother = choose a random live point
      # for supremum norm, in d dimensions
      v = draw d uniform random 
          numbers U(-R/2, R/2)
      candidate = mother + v

      # euclidean norm, in d dimensions
      v = draw d univariate Gaussian numbers
      v = normalize vector(v)
      u = draw uniform random number U(0, 1)
      candidate = mother + R * u**(1/d) * v
      # rejection
      m = count live_points with distance
                    less than R to candidate
      with probability 1/m, return candidate
    }
}
\end{lstlisting}

\par\end{centering}

\end{minipage}\hfill

\caption{\label{alg:draw_near}Pseudo-code for sampling a new point within
the sampling region defined by proximity within $R$ to a live point.
This can then be done in two ways, which are equivalent with regards
to the number of likelihood evaluations and properties of the generated
samples (see text). The second variant (right algorithm) is more elaborate,
and explained below.\protect \\
Here, the case of a Euclidean norm and the Supremum norm is illustrated.
In case of the Euclidean norm, each live point is surrounded by a
sphere of same radius (namely $R$). Sampling in the neighbourhood
of a point $\mathbf{p}$ can be done as follows: Drawing $d$ values
from a univariate Gaussian distribution, and normalising the resulting
vector yields a $d$-dimensional unit vector $\mathbf{\hat{v}}$ in
a random direction. Then, the length $r$ between 0 and \texttt{distance}
$R$ has to be chosen. Here, we have to keep in mind that higher dimensions
are less likely to generate a length close to 0. The correct approach
is to compute $r=R\times u^{1/d}$ with $u$ being a uniform random
number between 0 and 1. Finally, the new point is computed as $\mathbf{q}=r\cdot\mathbf{\hat{v}}+\mathbf{p}$.
\protect \\
For the supremum norm, the sampling is even easier. Computing $d$
uniform random numbers between $\pm R/2$ yields a vector $\mathbf{v}$.
The new point is then at $\mathbf{q}=\mathbf{v}+\mathbf{p}$.}
\end{algorithm*}

The \textsc{RadFriends} constrained sampling algorithm has to sample
a new live point subject to the constraint that it has a higher likelihood
value than $L_{\text{min}}$. It proceeds as laid out in the \texttt{draw\_constrained}
in Listing \ref{alg:RadFriends}. The \texttt{compute\_R} procedure
computes the aforementioned $R$, which is the largest distance to
a neighbour. Here a bootstrap-like procedure is employed to generate
a conservative estimate of $R$ by always leaving points out, and
ensuring they could be sampled. This distance $R$ is then used to
define the region around the live points to sample from.

The sampling procedure \texttt{draw\_near} can then be done in two
ways, which are equivalent with regards to the number of likelihood
evaluations and properties of the generated samples. Both are shown
in Algorithm~\ref{alg:draw_near}. The simpler method is to sample
a random point from the prior and check if it is within \texttt{distance}
of at least one live point. If not, the procedure is repeated. The
second method is to choose a random live point, and to generate a
random point that fulfils the distance criterion by construction (see
caption of Algorithm~\ref{alg:draw_near}). The so-generated point
must only be accepted with probability $1/m$, where $m$ is the number
of live points within distance $R$, to avoid preference to clustered
regions. The second method is more efficient than the first if the
remaining volume is small, as otherwise many points are rejected.

The remaining choice is which norm to use to define the distance.
Here I consider the Euclidean ($L_{2}$) norm $\left\Vert x\right\Vert $,
and the supremum ($L_{\infty}$) norm $\sup\left|x\right|$ (see Listing
\ref{alg:draw_near}). I term the variant of \textsc{RadFriends} that
uses the supremum norm \textsc{SupFriends}.

\subsection{Analysis of the emergent properties\label{sub:emerging}}

Figure \ref{fig:RadFriends-Contours} illustrates the behaviour of
the constructed sampling region under live points sampled from various
likelihood contours (green) in each column. The algorithm adapts its
sampling region (red and orange contours for the euclidean and supremum
norm respectively) to the existing points. Increasing the number of
live points tightens the sampling region. It can also be observed
that when one live point is far away from the others, the sampling
region is large, when they are close together, it tightens.

One curious choice in the algorithm is the number of bootstrap iterations
(given as 50). It was chosen as follows: The probability to not use
a specific live point in an iteration is 
\[
p_{1}=\left(1-\frac{1}{N}\right)^{N}\approx37\%\text{\,\,\text{for \ensuremath{N}>50}}.
\]
The probability to having used one particular point in \textit{every}
of the $m$ iterations, i.e. never having left it out, is 
\[
p_{L}=\left(1-p_{1}\right)^{m}.
\]
The probability of having used \textit{any} of the $N$ points in
\emph{every} iteration, is $N$ times higher. Here I neglect the subtraction
that this is the case for more than one point, which leads to the
upper-bound

\[
p_{L,all}<\left(1-p_{1}\right)^{m}\times N.
\]
This event should be rare, such that it should not be expected more
than once in the whole nested sampling run, e.g. with $10^{6}$ iterations.
For values of $N=100,\,1000,\,10000$, $p_{L,all}$ reaches the value
$10^{-6}$ at
\[
m=\frac{\ln\, p_{L,all}-\ln\, N}{\ln\,(1-p_{1})}=39.8,\,44.9,\,49.8.
\]
Thus, the conservative choice of 50 iterations is justified.

Figure \ref{fig:RadFriends-Contours} already demonstrates that this
algorithm can immediately handle multiple modes, as clustering of
points is an emergent feature. This yields efficient sampling iff
the region in between is excluded. When is this the case? Consider
a small cluster with $k$ points, well separated from the other live
points. It will be treated as a separate cluster if one of the members
is always selected in the bootstrapping rounds. Leaving out all $k$
points simultaneously has probability $p_{k,all}=p_{1}^{k}\times m$.
For $m=50$, and $k=10,\,20,\,40$, this probability is $p_{k,all}=0.5,\,0.005,\,5\times10^{-7}$.
In words, one can expect efficient sampling of the sub-cluster if
it contains more than 20 points. However, this means that for a problem
with e.g. $20$ well-separated modes, $20\times40=800$ live points
are needed to safely avoid the inefficient sampling between the modes.

\section{Shrinkage test results\label{sec:Evaluation}}

Now it is interesting to see whether the RADFRIENDS algorithm can
pass the shrinkage test constructed in Section \ref{sec:Shrinkage-Test}.
Additionally, I report the performance of a number of other algorithms,
namely plain rejection sampling, MULTINEST, MULTINEST with importance
nested sampling, and MCMC. For the constrained sampling using MCMC,
I employ a symmetric Gaussian proposal distribution of initial standard
deviation $0.1$ and test 10, 20 and 50 proposal steps. As the scales
shrink, an adaptive rule has to be used for the scale of the proposal
distribution. I use the update recipe described in \citet{sivia2006data}
of 

\[
\sigma'=\sigma\cdot\exp\left(\begin{cases}
+1/n_{\text{accepts}} & \text{if }n_{\text{accepts}}>n_{\text{rejects}}\\
-1/n_{\text{rejects}} & \text{if }n_{\text{accepts}}<n_{\text{rejects}}
\end{cases}\right)
\]
For comparison, I use another MCMC algorithm with a fixed Gaussian
proposal distribution of standard deviation $10^{-5}$, but 200 steps.

\begin{table*}
\begin{centering}
\begin{tabular}{l l | r r r r r }
Algorithm & dim & $p_{shrinkage}$ & iterations & evaluations & efficiency \\
\hline
rejection & 2 & 0.7324 & 32000 & 71092909 & 0.05\% \\ 
multinest & 2 & *{0.0474} & 80000 & 256411 & 31.20\% \\ 
radfriends & 2 & 0.9105 & 80000 & 132026 & 60.59\% \\ 
supfriends & 2 & 0.5321 & 80000 & 131505 & 60.83\% \\ 
mcmc-gauss-50-adapt & 2 & 0.1961 & 80000 & 4000000 & 2.00\% \\ 
mcmc-gauss-20-adapt & 2 & 0.1566 & 80000 & 1600000 & 5.00\% \\ 
mcmc-gauss-10-adapt & 2 & 0.0732 & 80000 & 800000 & 10.00\% \\ 
mcmc-gauss-scale-5  & 2 & *{0.0000} & 80000 & 16000000 & 0.50\% \\ 
\hline
rejection & 7 & 0.5707 & 32000 & 74035891 & 0.04\% \\ 
multinest & 7 & *{0.0000} & 80000 & 393575 & 20.33\% \\ 
radfriends & 7 & 0.2651 & 80000 & 2711519 & 2.95\% \\ 
supfriends & 7 & 0.0965 & 80000 & 3483200 & 2.30\% \\ 
mcmc-gauss-50-adapt & 7 & 0.3643 & 80000 & 4000000 & 2.00\% \\ 
mcmc-gauss-20-adapt & 7 & *{0.0273} & 80000 & 1600000 & 5.00\% \\ 
mcmc-gauss-10-adapt & 7 & *{0.0000} & 80000 & 800000 & 10.00\% \\ 
mcmc-gauss-scale-5  & 7 & *{0.0000} & 80000 & 16000000 & 0.50\% \\ 
\hline
rejection & 20 & 0.5183 & 32000 & 65401030 & 0.05\% \\ 
multinest & 20 & *{0.0000} & 32000 & 499209 & 6.41\% \\ 
radfriends & 20 & 0.2954 & 0.2954 & 32000 & 26129495 & 0.12\% \\ 
supfriends & 20 & 0.6573 & 32000 & 39067739 & 0.08\% \\ 
mcmc-gauss-50-adapt & 20 & 0.8785 & 32000 & 1600000 & 2.00\% \\ 
mcmc-gauss-20-adapt & 20 & 0.4475 & 32000 & 640000 & 5.00\% \\ 
mcmc-gauss-10-adapt & 20 & *{0.0000} & 32000 & 320000 & 10.00\% \\ 
mcmc-gauss-scale-5  & 7 & *{0.0000} & 80000 & 16000000 & 0.50\% \\ 
\hline

\end{tabular}
\par\end{centering}

\caption{\label{tab:Tests}Results of the shrinkage test using the hyper-pyramid
likelihood function. The p-value of the KS test indicates the expected
frequency of the result (values below 0.05 are indicated with a star).
In each algorithm, 400 live points were used. The rejection sampling
is run for fewer iterations as its efficiency drops rapidly. For exploration
with MCMC, the value indicates the number of proposal steps used (10,
20 or 50).}
\end{table*}

\begin{figure}
\begin{centering}
\includegraphics[bb=0bp 0bp 318bp 351bp,width=0.8\columnwidth]{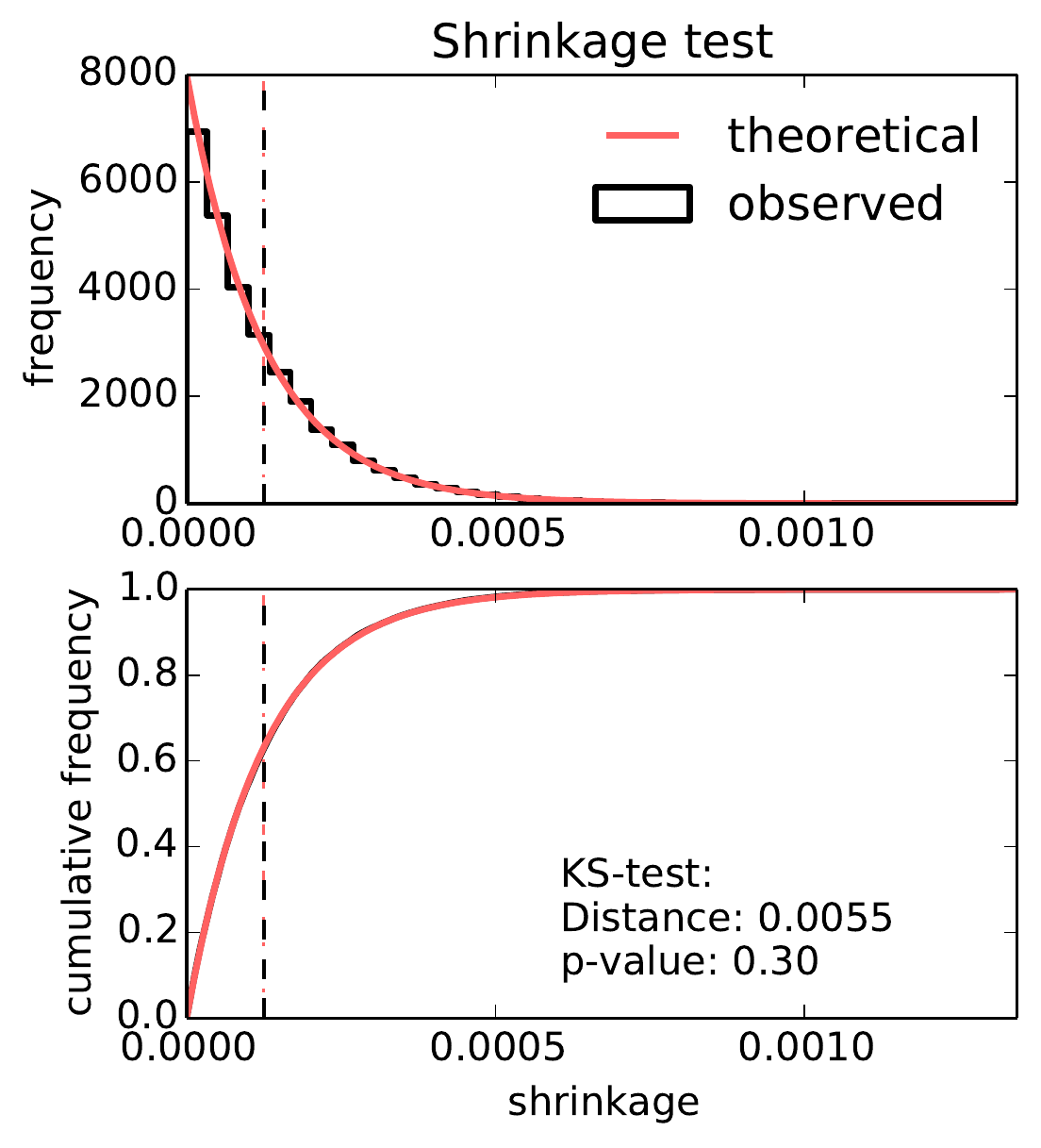}
\par\end{centering}

\caption{\label{fig:plot-supfriends}Same as Figure \ref{fig:plot-multinest}
but for the \textsc{SupFriends} algorithm in 20 dimensions. Here,
the distributions are in agreement.}
\end{figure}

The results are listed in Table \ref{tab:Tests}. The MCMC algorithm
with a tiny, fixed proposal (``mcmc-gauss-scale-5'') fails the Shrinkage
test as expected. It samples too close to the existing live points
(where it starts) and thus the shrinking is also incorrect. In contrast,
the MCMC proposal with an adaptive rule successfully passes the distance
distribution test. For the 7 and 20-dimensional case, the p-values
of either tests attain low values when using only 10 or 20 steps.
Although p-values can be cumbersome to interpret, it is sensible to
use at least 50 MCMC steps in the exploration, which yields an efficiency
of 2\%.

In 7 and 20 dimensions, the shrinkage distribution of the MULTINEST
algorithm shows deviations, as remarked before, and shown in Figure
\ref{fig:plot-multinest}. For comparison, the rejection sampling
and \textsc{RADFRIENDS} algorithm (shown in Figure \ref{fig:plot-supfriends})
yield the correct distribution.

Table \ref{tab:Tests} also shows that the MULTINEST algorithm is
highly efficient. In typical applications, the MULTINEST algorithm
uses one or up to two orders of magnitude fewer likelihood evaluations
than the \textsc{RADFRIENDS/SUPFRIENDS} algorithm.

\section{Test problems}

\begin{figure}
\begin{centering}
\includegraphics[width=1\columnwidth]{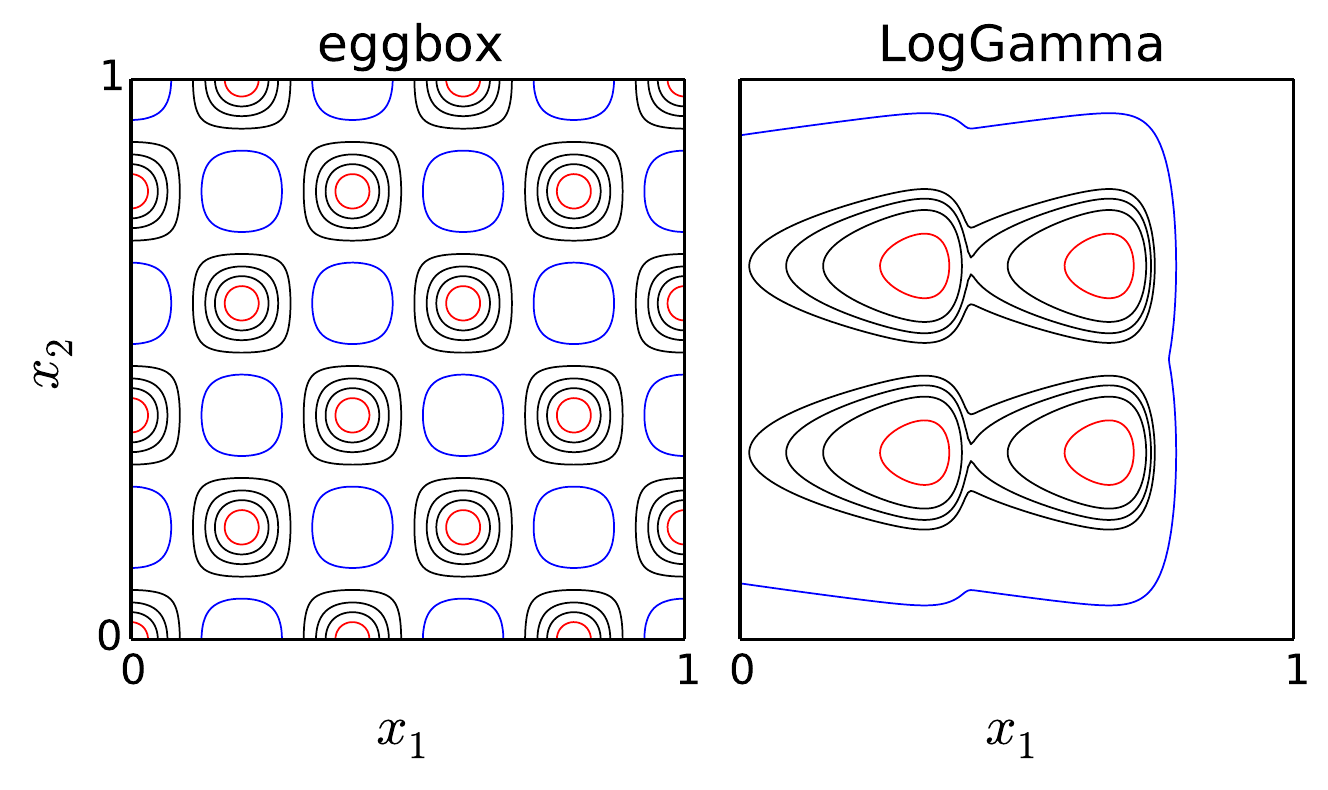}
\par\end{centering}

\caption{\label{fig:viz-problems}Visualisation of the considered problems
in the first two coordinates, using arbitrarily chosen contours (blue
lowest, red highest). Both the eggbox problem (\textit{left panel})
and the LogGamma problem (\textit{right panel}) show multi-modality.
For the latter, the contours are asymmetric. In higher dimensions,
the LogGamma problem is extended with independent Normal and LogGamma
distributions in alternation.}
\end{figure}
In this section, I analyse the correctness and efficiency of the RADFRIENDS\textsc{
}algorithm numerically. A number of common test integration problems
have been verified, however for brevity only two are presented here,
which expose the advantages and disadvantages best. For comparison,
I include results from using MULTINEST with and without Importance
Nested Sampling \citep{Feroz2013}. I run each algorithm 10 times,
and record the average integral value, $\hat{Z}$, the actual variance
of this estimator, $A^{2}$, and the average statistical uncertainty
reported, $C$. 

\newcommand{\problemshelltwod}{
multinest-nlive400 & 2 = & 0.7295 & 0.0727 & 0.0813 & 5960 \\
multinest-nlive400-INS & 2 = & 0.7686 & 0.0345 & 0.0207 & 6007 \\
radfriends-nlive400 & 3 (speed) & 0.7238 & 0.0677 & 0.4993 & 7538 \\
supfriends-nlive400 & 3 = & 0.7238 & 0.0677 & 0.4993 & 7538 \\
multinest-nlive1000 & 3 = & 0.7348 & 0.0404 & 0.0513 & 14887 \\
multinest-nlive1000-INS & 3 = & 0.7817 & 0.0458 & 0.0131 & 14947 \\
radfriends-nlive1000 & 3 = & 0.7729 & 0.0456 & 0.4998 & 13711 \\
supfriends-nlive1000 & 3 = & 0.7729 & 0.0456 & 0.4998 & 13711 \\
}

\newcommand{\problemshellonezerod}{
multinest-nlive400 & 1 & -11.9064 & 0.2794 & 0.1949 & 24902 \\
multinest-nlive400-INS & 1 = & -12.1421 & 0.0728 & 0.0230 & 25046 \\
multinest-nlive1000-INS & 2 (speed) & -12.0962 & 0.0131 & 0.0146 & 60724 \\
multinest-nlive1000 & 2 = & -11.8556 & 0.2782 & 0.1230 & 61446 \\
radfriends-nlive400 & 3 (speed) & -12.1540 & 0.1357 & 0.4996 & 651908 \\
supfriends-nlive400 & 3 = & -12.1540 & 0.1357 & 0.4996 & 651908 \\
radfriends-nlive1000 & 3 = & -12.0950 & 0.0928 & 0.4998 & 1123983 \\
supfriends-nlive1000 & 3 = & -12.0950 & 0.0928 & 0.4998 & 1123983 \\
cuba-Cuhre & 4 (accuracy) & -566.1907 & 554.0851 & 0.8825 & 7815 \\
cuba-Divonne & 5 (misleading) & -14.7354 & 2.7079 & 0.2775 & 13176 \\
cuba-Vegas & 6 (failed) & (failure) & (failure) & (failure) & 2002000 \\
cuba-Suave & 7 (misleading) & nan & nan & nan & 926770 \\
}

\newcommand{\problemeggbox}{
radfriends-nlive400 & 235.7985 & 0.1074 & 0.4995 & 388408 \\
supfriends-nlive400 & 235.7985 & 0.1074 & 0.4995 & 388408 \\
multinest-nlive400  & 235.9216 & 0.1053 & 0.1235 & 11077 \\
multinest-nlive400-INS & 235.9058 & 0.1008 & 0.0680 & 10595 \\
radfriends-nlive1000 & 235.8038 & 0.1045 & 0.4997 & 33736 \\
supfriends-nlive1000 & 235.8038 & 0.1045 & 0.4997 & 33736 \\
multinest-nlive1000  & 235.9082 & 0.0854 & 0.0782 & 26375 \\
multinest-nlive1000-INS & 235.8418 & 0.0418 & 0.0167 & 25601 \\
}

\newcommand{\problemfunneltwod}{
cuba-Vegas & 1 & -1.9205 & 0.3902 & 0.2499 & 1300 \\
cuba-Suave & 1 = & -1.9140 & 0.3818 & 0.2874 & 1400 \\
cuba-Divonne & 2 (speed) & -1.6160 & 0.0440 & 0.1375 & 2947 \\
multinest-nlive400-INS & 3 (speed) & -1.6269 & 0.1436 & 0.0632 & 12112 \\
multinest-nlive400 & 3 = & -1.8214 & 0.2376 & 0.1037 & 12178 \\
cuba-Cuhre & 3 = & -1.3874 & 0.2726 & 0.3957 & 21515 \\
multinest-nlive1000 & 4 (speed) & -1.5043 & 0.3108 & 0.0714 & 36279 \\
multinest-nlive1000-INS & 4 = & -1.4542 & 0.2802 & 0.0579 & 36531 \\
radfriends-nlive1000 & 5 (speed) & -1.3621 & 0.3702 & 1.2574 & 653852 \\
supfriends-nlive1000 & 5 = & -1.3621 & 0.3702 & 1.2574 & 653852 \\
radfriends-nlive400 & 6 (accuracy) & -1.3122 & 0.5211 & 1.3051 & 657413 \\
supfriends-nlive400 & 6 = & -1.3122 & 0.5211 & 1.3051 & 657413 \\
}

\newcommand{\problemloggammamultimodaltwod}{
radfriends-nlive400 & 0.1069 & 0.1233 & 0.4995 & 5258 \\
supfriends-nlive400 & 0.0984 & 0.1424 & 0.4993 & 5493 \\
multinest-nlive400 & 0.0058 & 0.0756 & 0.0777 & 5858 \\
multinest-nlive400-INS & 0.0348 & 0.0383 & 0.0165 & 5902 \\
radfriends-nlive1000 & 0.0866 & 0.0958 & 0.4998 & 11451 \\
supfriends-nlive1000 & 0.0812 & 0.1025 & 0.4998 & 11647 \\
multinest-nlive1000 & -0.0019 & 0.0590 & 0.0491 & 14366 \\
multinest-nlive1000-INS & 0.0399 & 0.0442 & 0.0104 & 14576 \\
}

\newcommand{\problemloggammamultimodalonezerod}{
radfriends-nlive400 & 0.0615 & 0.2322 & 0.2080 & 2251442 \\
supfriends-nlive400 & -0.0248 & 0.1025 & 0.2099 & 17136103 \\
multinest-nlive400 & 1.1740 & 1.1790 & 0.2028 & 71835 \\
multinest-nlive400-INS & 0.1978 & 0.2711 & 0.0963 & 71447 \\
radfriends-nlive1000 & 0.0179 & 0.1286 & 0.1316 & 3781340 \\
supfriends-nlive1000 & 0.0323 & 0.5400 & 0.3235 & 19763999 \\
multinest-nlive1000  & 0.9916 & 0.9941 & 0.1289 & 162111 \\
multinest-nlive1000-INS & 0.1224 & 0.1276 & 0.0372 & 165827 \\
}

\subsection{Eggbox problem}

The eggbox problem is adapted from \citet{Feroz2009}. It is only
two-dimensional, but contains 18 distinct peaks, posing extreme multi-modality.
The likelihood, visualised in Figure \ref{fig:viz-problems} (left
panel), can be defined on a unit square as

\[
\ln\, L=\left(2+\cos(5\pi\cdot x_{1})\cdot\cos(5\pi\cdot x_{2})\right)^{5}
\]

\begin{figure*}
\begin{centering}
\includegraphics[width=12cm]{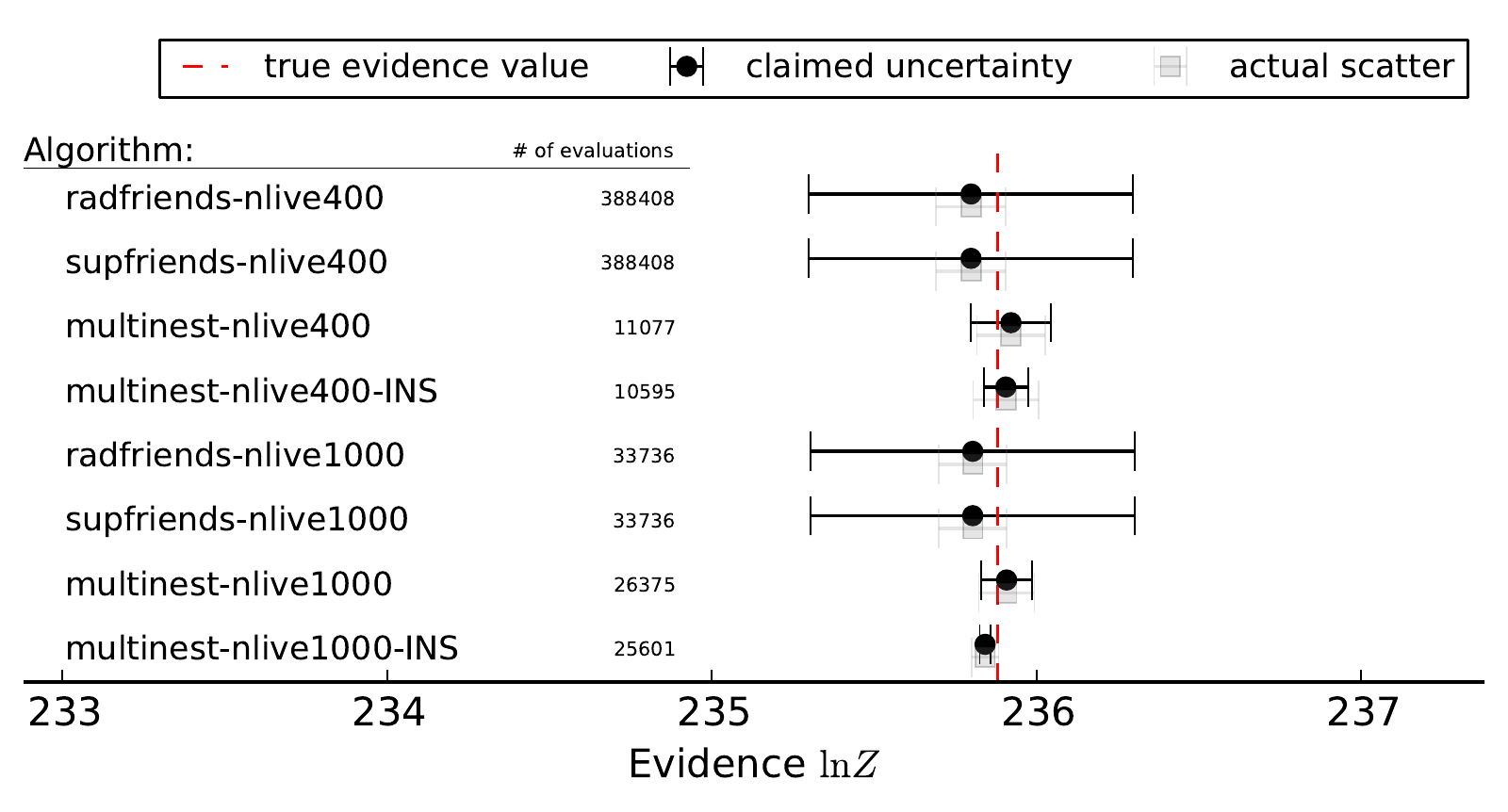}
\par\end{centering}

\caption{\label{tab:Results-eggbox}Performance results for the eggbox problem.
Each algorithm is listed with the mean $\ln\, Z$ indicated as a point.
For the uncertainties, the uncertainty of the estimate computed by
the algorithm is shown (black error bars). Grey error bars show the
estimators' actual scatter around the true value $\ln\, Z_{true}=235.88$
(vertical red dashed line). For each algorithm the total number of
likelihood function evaluations are listed. Here, all algorithms give
the correct answer. The \textsc{RadFriends} algorithm with 400 live
points yield a much lower efficiency than when using 1000 live points.
This is due to the many modes not being separated (see Section \ref{sub:emerging}).
The most efficient algorithm is MultiNest with 400 live points (1-2
orders of magnitudes faster than \textsc{RadFriends}).}
\end{figure*}

Results are shown in Figure \ref{tab:Results-eggbox}. Both MULTINEST
and\textsc{ }RADFRIENDS integrate this problem successfully. As appreciated
in Section \ref{sub:emerging},\textsc{ }RADFRIENDS can separate out
modes when a higher number of live points is used, making it more
efficient. MULTINEST uses the lowest number of likelihood evaluations.

\subsection{LogGamma problem}

This problem is adapted from \citet{Beaujean2013} and acknowledged
to be problematic by the MULTINEST authors \citep{Feroz2013}. A combination
of LogGamma and Gaussian distributions is considered, defining the
likelihood $L$ as 
\begin{eqnarray*}
g_{a} & \sim & \mathrm{LogGamma}\left(1,\,\frac{1}{3},\,\frac{1}{30}\right)\\
g_{b} & \sim & \mathrm{LogGamma}\left(1,\,\frac{2}{3},\,\frac{1}{30}\right)\\
n_{c} & \sim & \mathrm{Normal}\left(\frac{1}{3},\,\frac{1}{30}\right)\\
n_{d} & \sim & \mathrm{Normal}\left(\frac{2}{3},\,\frac{1}{30}\right)\\
d_{i} & \sim & \mathrm{LogGamma}\left(1,\,\frac{2}{3},\,\frac{1}{30}\right)\,\text{\,\,\ if\,\,\,}3\leq i\leq\frac{d+2}{2}\\
d_{i} & \sim & \mathrm{Normal}\left(\frac{2}{3},\,\frac{1}{30}\right)\,\text{\,\,\,\,\,\,\,\,\,\,\,\,\,\,\,\,\,\,\,\,\ if\,\,\,}\frac{d+2}{2}<i\\
L_{1} & = & \frac{1}{2}\left(g_{a}(x_{1})+g_{b}(x_{1})\right)\\
L_{2} & = & \frac{1}{2}\left(n_{c}(x_{2})+n_{d}(x_{2})\right)\\
L & = & L_{1}\times L_{2}\times\prod_{i=3}^{d}d_{i}(x_{i}).
\end{eqnarray*}
The dimensionality of the problem is donated by $d$. We consider
the cases of $d=2$ and $d=10$ here. This problem combines well-separated
peaks with asymmetric heavy-tailed distributions, as shown in Figure
\ref{fig:viz-problems}. The true integral value is $\ln\, Z_{\text{true}}=0$.

\begin{figure*}
\begin{centering}
\includegraphics[width=12cm]{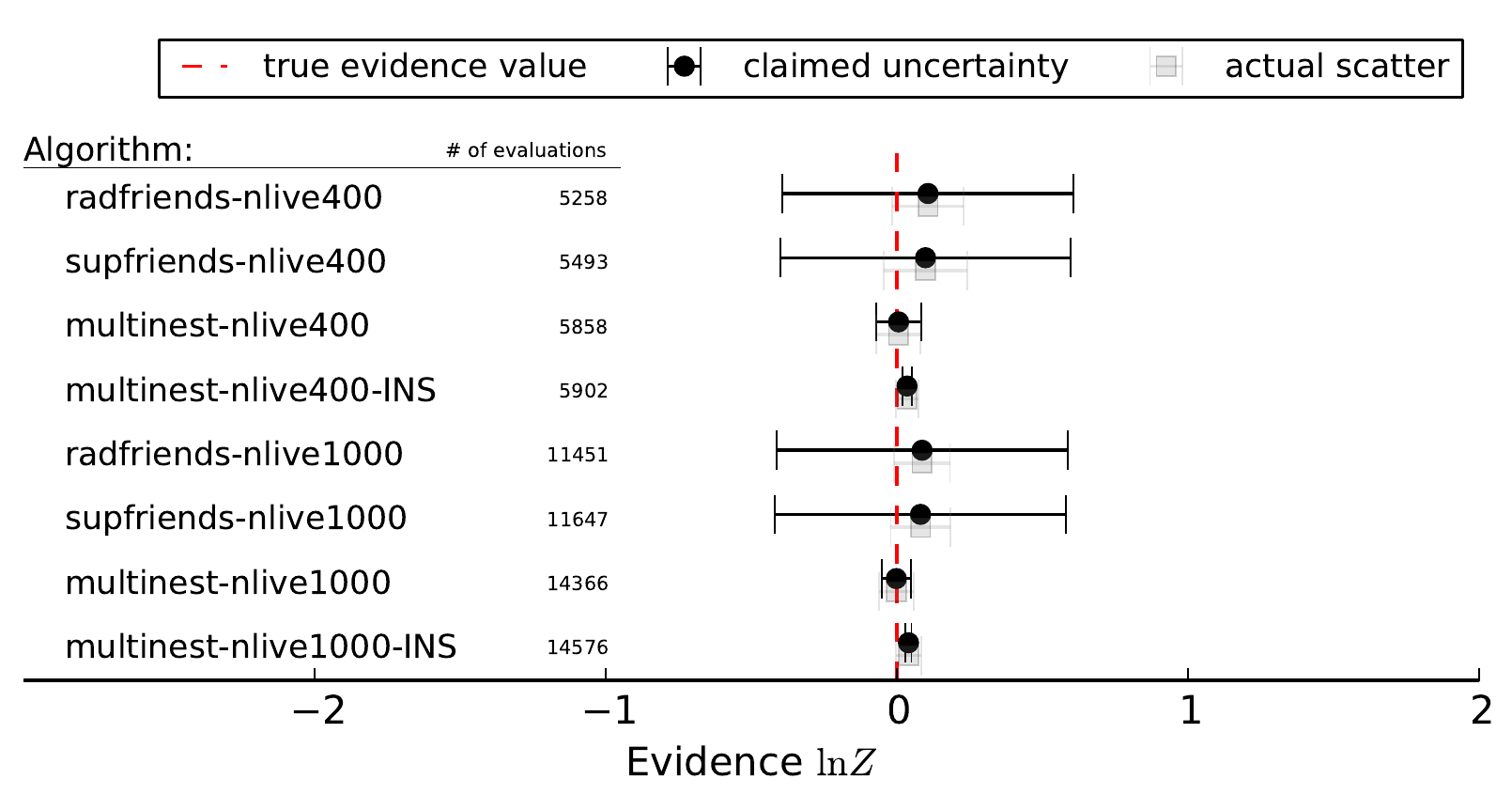}
\par\end{centering}

\caption{\label{tab:LogGamma2}Performance results for the LogGamma problem
in 2 dimensions. Each algorithm is listed with the mean $\ln\, Z$
indicated as a point. For the uncertainties, the uncertainty of the
estimate computed by the algorithm is shown (black error bars). Grey
error bars show the estimators' actual scatter around the true value
$\ln\, Z_{true}=0$ (vertical red dashed line). For each algorithm
the total number of likelihood function evaluations are listed. All
algorithms give correct results. However, MULTINEST with Importance
Nested Sampling claims a much smaller uncertainty than actually achieved,
excluding the true value.}
\end{figure*}

\begin{figure*}
\begin{centering}
\includegraphics[width=12cm]{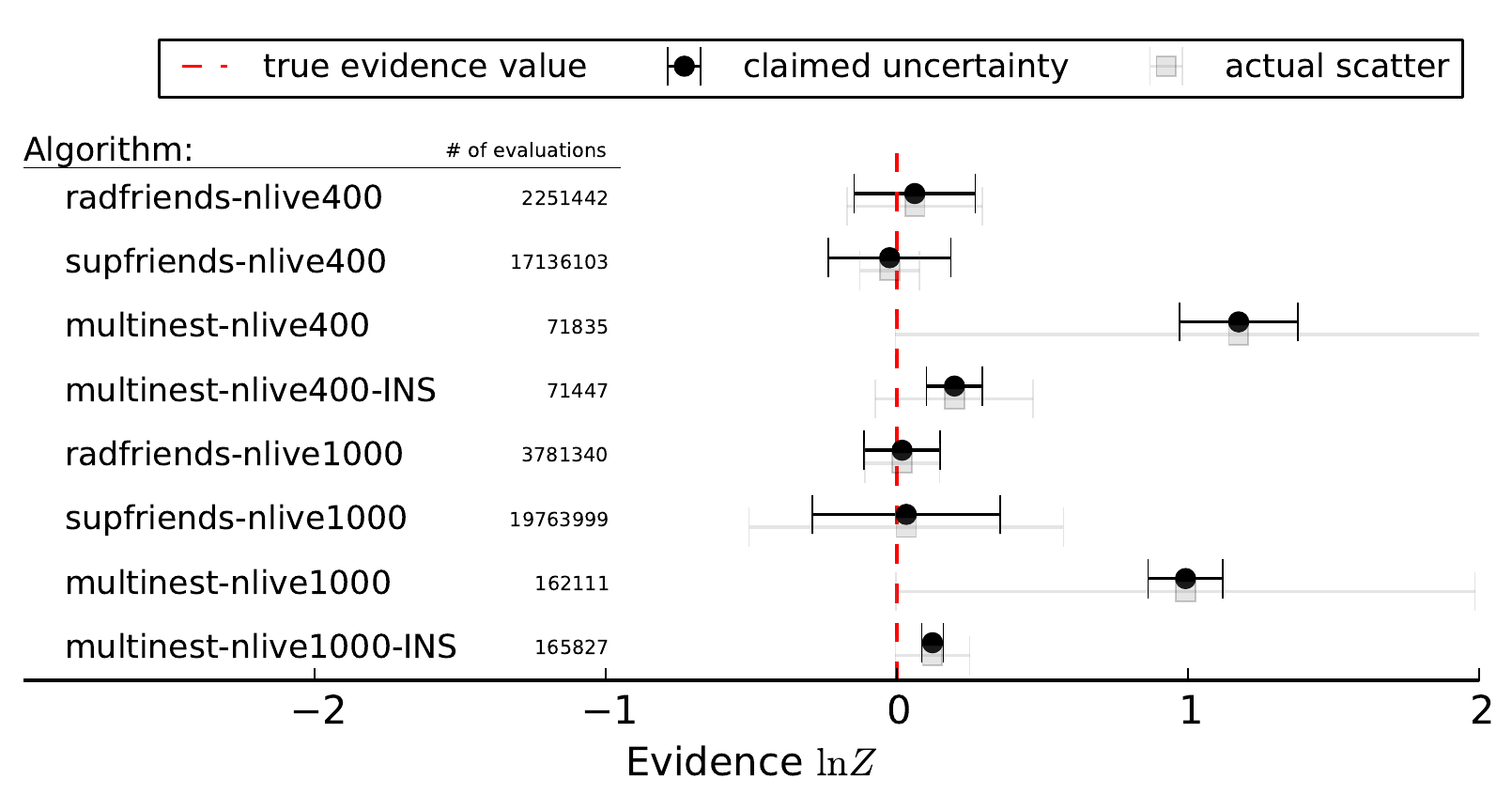}
\par\end{centering}

\caption{\label{tab:LogGamma10}As Table \ref{tab:LogGamma2}, but in 10 dimensions.
Here, MULTINEST overestimates the evidence. Enabling Importance Nested
Sampling reduces the over-estimate. The \textsc{RadFriends/SupFriends
}algorithms yield the correct results, and do not show any bias. Using
the supremum norm requires an order of magnitude more evaluations.}
\end{figure*}

The results are shown in Figures \ref{tab:LogGamma2} for the two-dimensional
case and Figure \ref{tab:LogGamma10} for ten dimensions. The two-dimensional
problem can be solved correctly (i.e. within the constraints) by all
algorithms. However, the Importance Nested Sampling of MULTINEST claims
a higher accuracy (by a factor of $\sim5$) than actually achieved.
This effect has been noted before in \citet{Feroz2013}.

The 10-dimensional problem demonstrates what happens when the algorithms
begin to break. Without Importance Nested Sampling, the computation
terminates, but the found integral value is over-estimated. With Importance
Nested Sampling enabled, MULTINEST mitigates the overestimation to
sufficient degree. Both\textsc{ }RADFRIENDS and SUPFRIENDS compute
the evidence correctly, which shows that this problem can be solved
by standard nested sampling. SUPFRIENDS requires one magnitude more
evaluations than RADFRIENDS, which indicates that the choice of the
norm has a strong influence for problems of higher dimensionality.

\section{Conclusions}

We have presented a brief overview of algorithms for sampling under
a constrained prior, which are a key ingredient in nested sampling,
and employed to compute integrals for high-dimensional model comparison.
We studied the sources of errors in such algorithms and devise a test
to uncover sampling errors.

The Shrinkage Test uncovers algorithms that violate the expectation
of nested sampling in how the prior volume shrinks. Such problematic
algorithms accelerate the shrinking, leaving out relevant parameter
space, which leads to incorrect computation of the integral.

Although the Shrinkage Test is limited to geometrically well-understood
likelihood functions with geometrically simple contours (such as Gaussian
likelihoods, or the hyper-pyramid used here), it can be used to verify
the correctness on high-dimensional problems, multi-modal likelihoods,
and shapes of multiple scale lengths. Thus, it capable of simulating
a wide range of situations that occur in practise.

We apply the Shrinkage Test to the popular MULTINEST algorithm, and
find that it fails in the 7 and 20-dimensional cases. This indicates
that in the studied case, relevant prior volume is left out. This
type of error may also the source for not integrating the LogGamma
problem correctly.

We then present an algorithm termed RADFRIENDS, which is constructed
to be robust against this type of problem. Studying the properties,
we find that RADFRIENDS
\begin{enumerate}
\item passes the Shrinkage Test,
\item solves the LogGamma problem and others correctly, and
\item can handle multi-modal problems and peculiar shapes without tuning
parameters or additional input information.
\end{enumerate}
However, this algorithm is one or two orders of magnitudes less efficient
than MULTINEST by number of likelihood evaluations. This algorithms
suffers from the curse of dimensionality and is thus not useful for
$>10$ dimensions, save for verifying test problems with fast-to-compute
likelihoods. For low-dimensional problems, it can, however, compete
with MULTINEST.

The proposed algorithm is simple to implement, and can be understood
analytically. We propose its use as a safe, easy-to-implement baseline
algorithm for low-dimensional problems.

In a similar spirit, the method of \citet{Mukherjee2006} and the
MULTINEST algorithm could be made more robust. We suggest leaving
a fraction of the live points out when constructing the ellipsoids.
The ellipsoids should then be expanded to such a degree that the left-out
live points are included. This can be done a few times to obtain a
robust ellipsoid expansion factor, on-line.

\section{Future Work}

The \emph{region sampling} type of constrained sampling algorithms,
which constructs a sampling region from the live points, requires
further study, especially in the high-dimensional regime. For instance,
machine learning algorithms, such as Support Vector Machines, may
be useful to learn the border between live points and already discarded
points. Improvements and further studies of the simple \textsc{RADFRIENDS}
algorithm are also left to future work. For example, applying Importance
Nested Sampling \citep{Cameron2013} in RADFRIENDS is directly analogous
to how it was developed for MULTINEST in \citet{Feroz2013}. The study
of the impact of the distance measure, and alternative norms may also
be useful for higher dimensional problems.

The option of combining \emph{region sampling} and \emph{local step}
methods into hybrid algorithms should be explored to combine their
respective power. For instance, the permissible region from \textsc{RADFRIENDS}
may be used as a restrict the proposal distribution of Markov Chain
Monte Carlo, or its hyper-spheres may be used as reflection surfaces
for Galilean Monte Carlo. The scale-size of the region ($R$) can
also be used to tune the step size. Such a \textsc{RadFriends}/MCMC
hybrid method written in C, named \textsc{UltraNest, }is available
at \url{http://johannesbuchner.github.io/nested-sampling/UltraNest/}.
A framework for developing and testing nested sampling algorithms
in Python is available at \url{http://johannesbuchner.github.io/nested-sampling/},
for which we welcome contributions. A reference implementation of
\textsc{RADFRIENDS} can also be found there.

\acknowledgement{I would like to thank Frederik Beaujean and Udo
von Toussaint for reading the initial manuscript. I acknowledge funding
through a doctoral stipend by the Max Planck Society. This manuscript
has greatly benefited from the comments of the two anonymous referees,
whom I would also like to thank. I acknowledge financial support through
a Max Planck society stipend.}

\bibliographystyle{apalike2}
\bibliography{/mnt/data/daten/PhD/research/agn/agn}

\end{document}